\documentclass[manuscript]{emulateapj}
\usepackage{natbib}
\usepackage{amsmath}
\usepackage{color}

\newcommand{\afe}{[$\alpha$/Fe]}
\newcommand{\sgm}{$\sigma$}
\newcommand{\tSF}{$\tau_{{\rm SF}}$} 
\newcommand{\lgMF}{$\log ($Mgb$/\langle {\rm Fe} \rangle)$}

\newcommand{\kms}{km~s$^{-1}$}

\begin{document}

\title{ The Influence of Environment on the Chemical Evolution in Low-mass Galaxies }

\author{Yiqing Liu\altaffilmark{1,2}, Luis C. Ho\altaffilmark{2,1}, Eric Peng\altaffilmark{1,2}} 

\altaffiltext{1}{Department of Astronomy, Peking University, Beijing 100871, China. Email: yiqing.liu@pku.edu.cn}
\altaffiltext{2}{Kavli Institute for Astronomy and Astrophysics, Peking University, Beijing 100871, China}


\begin{abstract}
The mean alpha-to-iron abundance ratio (\afe) of galaxies is sensitive to the chemical evolution processes at early time, and it is an indicator of star formation timescale (\tSF). Although the physical reason remains ambiguous, there is a tight relation between \afe\ and stellar velocity dispersion (\sgm) among massive early-type galaxies (ETGs). However, no work has shown convincing results as to how this relation behaves at low masses.  We assemble 15 data sets from the literature and build a large sample that includes 192 nearby low-mass ($18<\sigma<80$~\kms) ETGs.  We find that the \afe-\sgm\ relation generally holds for low-mass ETGs, except in extreme environments. Specifically, in normal galaxy cluster environments, the \afe-\sgm\ relation and its intrinsic scatter are, within uncertainties, similar for low-mass and high-mass ETGs.  However, in the most massive relaxed galaxy cluster in our sample, the zero point of the relation is higher and the intrinsic scatter is significantly larger.  By contrast, in galaxy groups the zero point of the relation offsets in the opposite direction, again with substantial intrinsic scatter.  The elevated  \afe\ of low-mass ETGs in the densest environments suggests that their star formation was quenched earlier than in high-mass ETGs. For the low-mass ETGs in the lowest density environments, we suggest that their more extended star formation histories suppressed their average \afe.  The large scatter in \afe\ may reflect stochasticity in the chemical evolution of low-mass galaxies. 
\end{abstract}


\section{Introduction}

Most early-type galaxies (ETGs) stopped forming stars long ago.  They 
provide a fossil record of the star formation histories (SFHs), quenching 
scenarios, and mass assembly in the early Universe. A powerful tracer of 
these early processes is the alpha-to-iron abundance ratio (\afe), which is an 
indicator of star formation timescale (\tSF).  In the first $\sim$0.1~Gyr 
after star formation begins, enrichment of the interstellar medium is 
dominated by Type ~II supernovae (SNe~II), which return ejecta with a 
relatively high \afe\ to the interstellar medium. Afterwards, Type~Ia 
supernovae (SNe~Ia) begin to contribute Fe-rich ejecta, and the \afe\ of the 
entire stellar system quickly decreases as stars continue to form from this 
Fe-enriched gas. Over time, the system reaches an equilibrium value close to 
the the solar value, \afe $\approx 0$ (Haywood et al. 2013).  Galaxies with 
shorter \tSF\ should reach higher \afe; \afe\ is sensitive to \tSF\ when \tSF\ 
is relatively short. 

Massive ETGs ($\sigma\gtrsim80$~\kms) obey a tight empirical relation between 
\afe\ and central stellar velocity dispersion (\sgm), such that galaxies with 
larger \sgm\ have higher \afe.  The slope and zero point of the  \afe-\sgm\ 
relation do not vary much with environment (e.g., Thomas et al. 2005; McDermid 
et al. 2015).  This relation indicates that more massive ETGs have shorter 
\tSF\ and, on average, quenched earlier. One possible explanation is 
feedback from active galactic nuclei (AGNs), as more massive ETGs host more 
massive central black holes (Kormendy \& Ho 2013).  The expected strong 
quenching by AGN feedback processes (e.g., King 2003) can affect the  
\afe-\sgm\ relation, according to the cosmological simulations of Segers 
et al. (2016).

Alternatively, the variation of the initial stellar mass function (IMF), 
fraction of SN~Ia binaries, the delay time distribution of SN~Ia, and stellar 
yields can also account for the positive correlation between \sgm\ and \afe\ 
(e.g., Arrigoni et al. 2010; Yates et al. 2013). However, the effects of 
these parameters on the \afe-\sgm\ relation are highly degenerate, and there 
is no consensus as to which dominates.  It is also possible that some
massive ETGs are remnants of wet major mergers, and that their high \afe\ was
set by the most recent starburst.  However, the frequency of major mergers is 
not high (e.g., Lotz et al. 2011), and this scenario probably cannot fully account for the \afe-\sgm\ relation. 

If quenching processes determine the \afe-\sgm\ relation of massive ETGs, then 
how this relation behaves at low mass is interesting. While the quenching 
mechanisms of massive galaxies are mass-dependent, low-mass galaxies are 
mostly quenched by environmental processes (Peng et al. 2010), such as the ram 
pressure stripping (Gunn \& Gott 1972), harassment (Moore et al. 1996), tidal 
stirring (Mayer et al. 2001), and starvation (Peng et al. 2015).  Even if 
low-mass galaxies are not quenched by environment, feedback from SN explosions 
and stellar winds are still distinct from the quenching mechanisms of massive 
galaxies (e.g., Hopkins et al. 2011; Forbes et al. 2016). 

Besides quenching, the chemical evolution histories of galaxies also depend on 
their mass.  Peng \& Maiolino (2014) pointed out that for star-forming galaxies, 
the key parameter that controls galactic evolution is the timescale required 
to reach equilibrium. While most massive galaxies reach equilibrium in their 
lifetime, low-mass ($M_*\lesssim10^9~M_\odot$), gas-rich galaxies may not 
within a Hubble time.  As a consequence of their non-equilibrium states, 
stochastic processes in low-mass galaxies, including inflow, outflow, and star 
formation, may be imprinted in their stellar population when they are 
quenched.  For instance, Lee et al. (2009) found a systematically lower 
H$\alpha$/FUV flux ratio with declining luminosity among low-mass, star-forming 
galaxies, which might be due to the stochastic appearance of high-mass stars 
at low star formation rates (SFRs; e.g., Fumagalli et al. 2011). 

Therefore, it is important to investigate the \afe-\sgm\ relation of low-mass 
ETGs. Unfortunately, no work has studied it using a sample of significant size, 
possibly due to their low surface brightness.  Previous studies all 
had relatively large uncertainties.  Sansom \& Northeast (2008), Smith et al. 
(2009), and Annibali et al. (2011) obtained different slopes for this relation.
Using a sample of low-mass ETGs from the Virgo Cluster, Liu et al. (2016) 
found a \afe-\sgm\ relation with larger scatter and, based on a correlation 
between \afe\ and distance from the cluster center, concluded that low-mass 
ETGs are more governed by environment instead of mass.  However, Vargas et al. 
(2014), from a sample of Local Group dwarf spheroidal galaxies (dSphs), 
claimed a large scatter without any environmental dependence.

To study the early baryonic processes in low-mass galaxies, we collect
literature data for a relatively large sample to systematically study the 
\afe-\sgm\ relation of low-mass ETGs. We are interested in not only the form 
of the relation, but also its intrinsic scatter, which provides information 
about the early SFHs of low-mass galaxies. \\

\section{Data}
\label{data}

We assemble from the literature a sample of 708 ETGs, which spans a wide range 
in mass and environment, with measurements of both \afe\ and \sgm.  In order 
to make the sample as homogeneous as possible and allow for inter-comparison, 
we only collect \afe\ measurements derived from single stellar population 
models, using Lick index measurements of integrated-light spectra. As a 
result, galaxies as faint as the Local Group dSphs are excluded from our 
sample, and no measurement is from high redshift.  Our final sample has 
\sgm\ ranging from 18 to 360~\kms, among them 192 low-mass ETGs with \sgm\ 
$<\, 80$~\kms. 

The \afe\ values of most of the data sets used in our analysis were derived 
following the methodology of Thomas et al. (2003) or Schiavon (2007).  To 
eliminate the systematic offsets between these two models, we apply the 
following correction from Smith et al. (2009) to adjust the values of Schiavon 
(2007; Sch) to the scale of Thomas et al. (2003; TMB): 

\begin{equation}
  [\alpha/{\rm Fe}]^{\rm TMB} = 0.99[{\rm Mg/Fe}]^{\rm Sch}  + 0.06[{\rm Fe/H}]^{\rm Sch} + 0.02 \\
 \label{E2M_afe}
\end{equation}
%
\noindent
The models of Proctor \& Sansom (2002) and Sansom \& Northeast (2008) are the 
same as that of Thomas et al. (2003), except that they did not account for the 
\afe\ bias in the solar neighborhood; this, however, has negligible effects on 
our results. The models of  Annibali et al. (2007, 2011) are consistent with 
those of Thomas et al. (2003).  A more detail discussion is given in 
Section~\ref{R_index}. 

\begin{figure*}
\epsscale{1.25}
\plotone{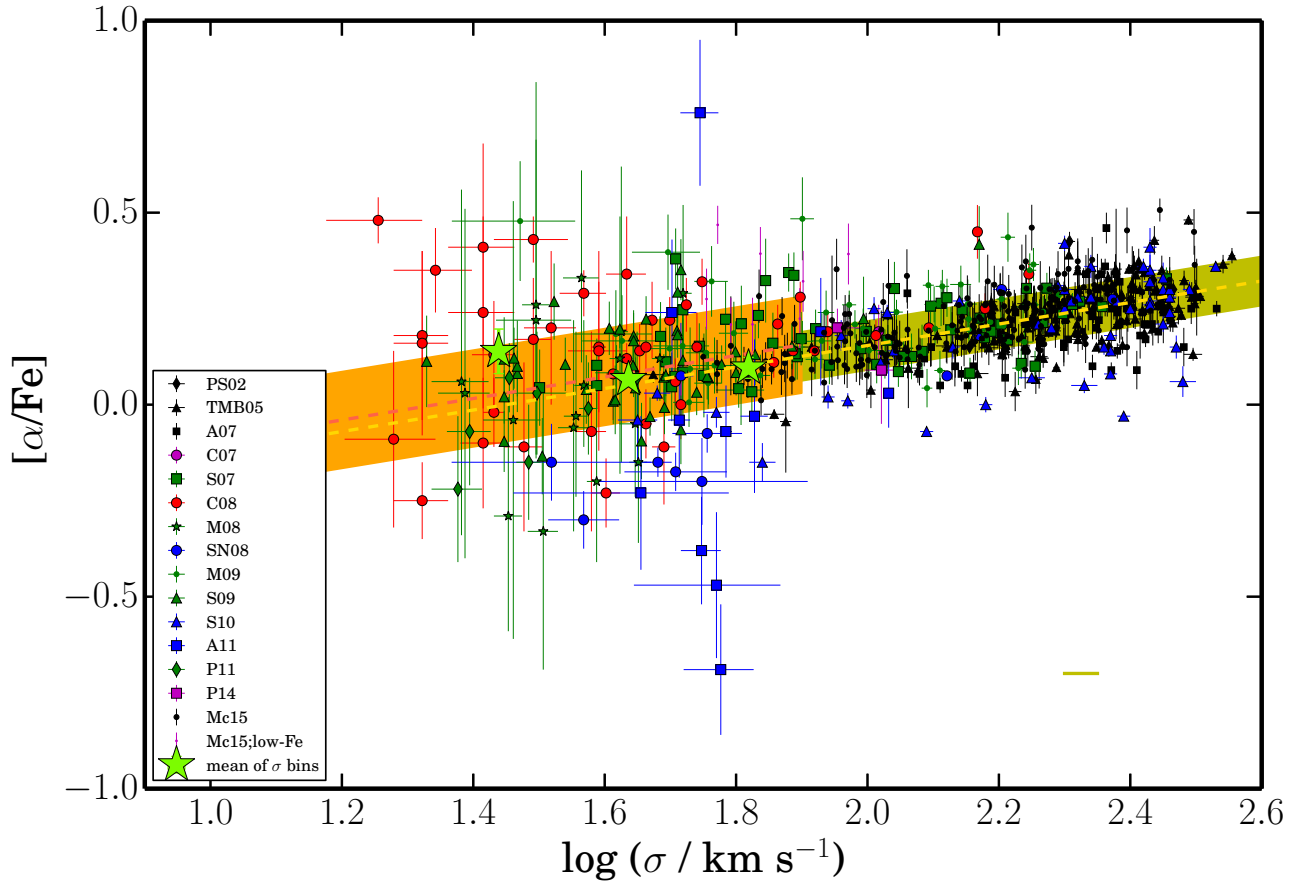}
\caption{ The relation between \afe\ derived from single stellar population 
models and \sgm\ of ETGs.  Different colors and symbols represent different 
subsamples, whose references are listed in the legend: Proctor \& Sansom (2002; PS02), Thomas et al. (2005; TMB05), Annibali et al. (2007; A07), Chilingarian et al. (2007; C07), Smith et al. (2007; S07), Chilingarian et al. (2008; C08), Michielsen et al. (2008; M08), Sansom \& Northeast (2008; SN08), Matkovi{\'c} et al. (2009; M09), Smith et al.  (2009; S09), Spolaor et al.  (2010; S10), Annibali et al. (2011; A11), Paudel et al.  (2011; P11), Paudel et al. (2014; P14), and McDermid et al. (2015; Mc15). The \sgm\ measurements of the Michielsen et al. (2008) and Paudel et al. (2011) sources are from Toloba et al. (2014).  
The galaxies in our sample are divided into two groups by $\log \sigma=1.9$. 
The green stars in the low-mass range show the weighted mean of three bins 
of \sgm, which are $\log \sigma = 1.2 - 1.5$, $1.5 - 1.7$, and $1.7 - 1.9$. 
The yellow dashed line and band display the relation and intrinsic scatter 
fitted by the high-\sgm\ group. In the low-\sgm\ range, they are displayed by 
the orange dashed line and band, and the 
 slope is fixed to be the same as that of the high-\sgm\ group in the fitting. 
 The potential offset of \sgm\ between different subsamples is shown at the 
 bottom-right corner; this has negligible effects on our results. \\
 \label{afeSgm} }
\end{figure*}

Our data sets were acquired using a range of aperture sizes, although most 
were between 1/8 to 1 galaxy effective radius ($R_e$). The more massive 
galaxies were measured from more central regions, whereas low-mass objects 
tend to cover relatively larger area.  We note, however, that ETGs generally 
have very weak radial gradients in \afe\ (e.g., Spolaor et al. 2010), and thus 
any systematic offsets in \afe\ from different apertures should be small.
As for velocity dispersions, Cappellari et al. (2013a,b) provide \sgm\ 
measurements at both $R_e$/8 and $R_e$ of 260 massive ETGs.  The mean absolute 
offset between these two apertures is 0.05~dex. As illustrated in 
Figure~\ref{afeSgm} (yellow line in the bottom-right corner), this level of 
potential systematic offset cannot strongly affect the global \afe-\sgm\ 
relation. \\

\section{Results}
\label{result}

\begin{figure*}
\epsscale{1.25}
\plotone{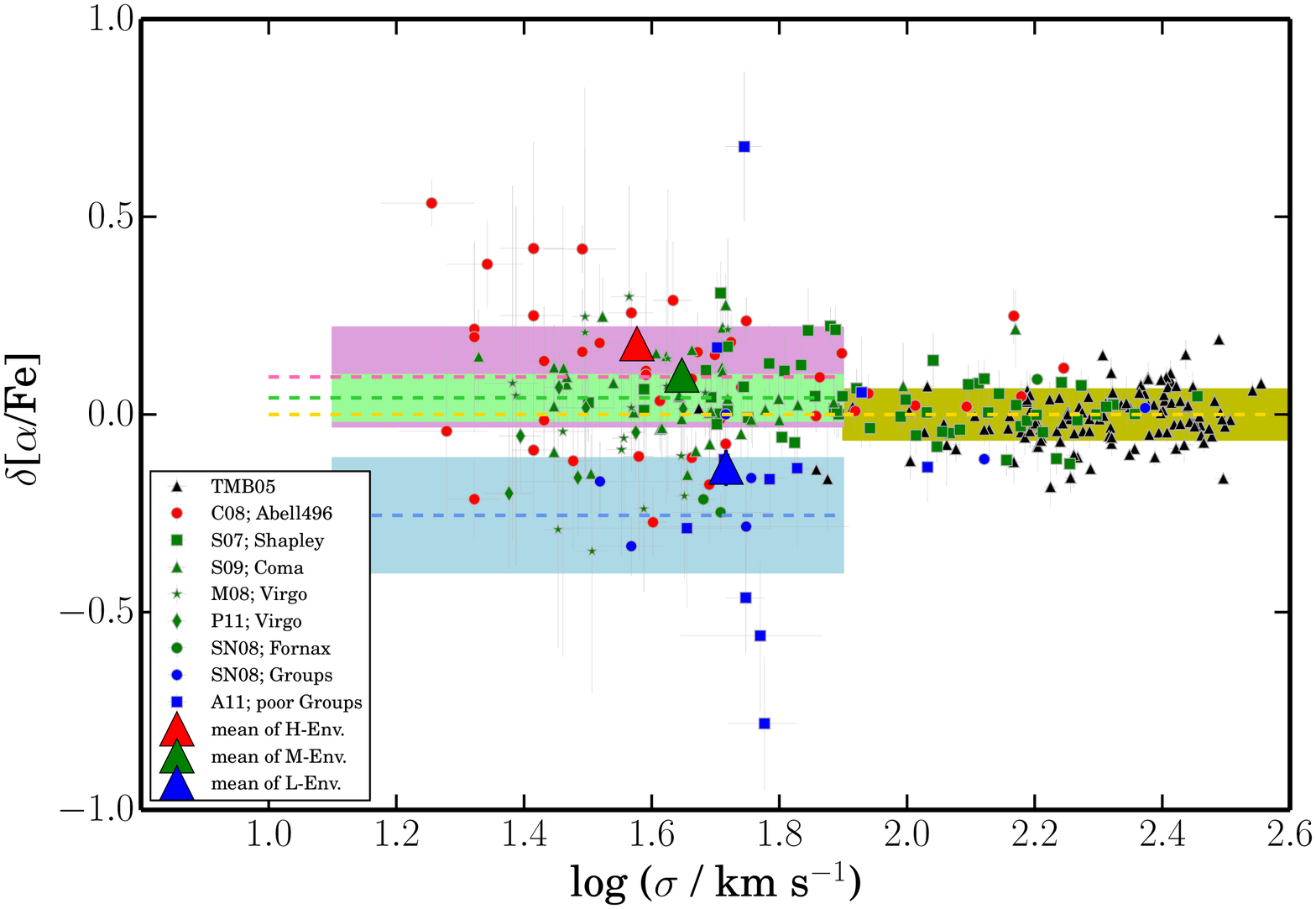}
\caption{ 
 The distribution of \afe\ residuals with \sgm\ and environment. The residuals of $\delta$[$\alpha$/Fe] are calculated 
 from the relation fitted by the high-\sgm\ group. Except for the objects from 
 the sample of TMB05 (black triangles), which is included for reference, only the samples 
 that contain low-\sgm\ galaxies and have environmental information are plotted. 
 The samples with the highest and lowest density environments are color-coded 
 red and blue, respectively; those located in intermediate-density environments are shown in green. The corresponding colored triangles 
 represent the weighted mean \afe\ and mean \sgm\ of the three environmental subsamples.  Dashed lines give the zero point offsets with respect to the high-\sgm\ group (yellow dashed line), and colored bands denote their respective 
 intrinsic scatter. \\
 \label{DafeEnv} }
\end{figure*}

\subsection{The \afe-\sgm\ Relation} 
\label{R_aMR}

Figure~\ref{afeSgm} displays the \afe-\sgm\ diagram of ETGs across a wide mass 
range, using the entire sample we collected. Different subsamples are plotted 
in different colors and symbols. 
Whereas the upper end of the distribution clearly obeys the well-known tight 
correlation, we see that the scatter notably increases toward the lower end.  
The transition occurs at $\log \sigma \approx 1.9$ ($\sigma=80$~\kms), which we 
use as the dividing line between the high-\sgm\ (high-mass) and low-\sgm\ 
(low-mass) groups.  

We begin by characterizing the \afe-\sgm\ relation for the high-\sgm\ group.  
To fit the linear relation and calculate their intrinsic scatter, we use 
{\tt MPFITEXY}, which is the {\tt FITEXY} estimator of Press et al. (1992), as
modified by Tremaine et al. (2002). We allow for asymmetric errors. To 
estimate the uncertainties caused by the potentially biased sampling, we 
perform 5000 bootstrap iterations on each fit to calculate the errors of our 
fitting parameters. The results, listed in Table~\ref{fitPar} and illustrated 
in Figure~\ref{afeSgm} (yellow dashed line and band), agree well with previous 
work (e.g., Thomas et al. 2005; McDermid et al. 2015). The intrinsic 
scatter of the \afe-\sgm\ relation for the high-\sgm\ group is very small, only 
$\epsilon_0 = (0.064\pm0.004)$ dex.

\begin{table}
\begin{center}
\caption{ The \afe-\sgm\ relation for different subsamples  
 \label{fitPar}}
\begin{tabular}{lccc}
\tableline\tableline
Group & $a$ & $b$ & $\epsilon_0$ \\
\tableline\tableline
High-\sgm\ & $0.280\pm0.023$ & $-0.406\pm0.051$ & $0.064\pm0.004$ \\
\tableline\tableline
Low-\sgm \\
\tableline\tableline
Total & $0.280$\footnotemark[1] & $-0.376\pm0.011$ & $0.126\pm0.014$ \\
H-Env & $0.280$\footnotemark[1] & $-0.311\pm0.030$ & $0.126\pm0.032$ \\
M-Env & $0.280$\footnotemark[1] & $-0.364\pm0.012$ & $0.058\pm0.015$ \\
L-Env & $0.280$\footnotemark[1] & $-0.661\pm0.051$ & $0.145\pm0.064$ \\
\tableline\tableline
\end{tabular}
\tablecomments{ \afe\ = $a \log \sigma + b$.  The intrinsic scatter is 
$\epsilon_0$.
H/M/L-Env represent the groups of low-mass 
ETGs with different levels of environmental density. \\
\footnotetext[1]{No error measurement, because the slope is fixed.} 
}
\end{center}
\end{table}

The large measurements uncertainties for the low-mass ETGs prevent us from 
performing an unconstrained fit for these objects.  To first obtain a 
qualitative assessment, we divide the low-mass group into three bins of $\log \sigma$: $1.2-1.5$, $1.5-1.7$, and $1.7-1.9$. In Figure~\ref{afeSgm}, the 
green stars show the weighted average values of these bins. The two bins 
with higher \sgm\ basically follow the extrapolation of the relation of the 
high-mass group.  Fixing the slope to that of the high-mass end ($a = 0.280$), 
we find a statistically similar zero point but an intrinsic scatter 
[$\epsilon_0 = (0.126\pm0.014)$ dex] two times larger than that of the 
high-mass objects (Table~\ref{fitPar}; orange dashed line and band in 
Figure~\ref{afeSgm}). \\

\subsection{The Environmental Dependence of the \afe-\sgm\ Relation} 
\label{R_env}

The large intrinsic scatter of the low-mass objects, however, is not 
contributed by the entire sample.  Most of the objects with the highest values 
of \afe\ are from the study of Chilingarian et al. (2008), which targeted the 
galaxy cluster Abell~496, the densest environment in our compilation.  At the 
same time, the galaxies with the lowest \afe\ are from low-density,
poor groups. This motivates us to further test the role of 
environment in driving the large scatter of the \afe-\sgm\ relation in the 
low-mass regime.  We divide our low-mass sample into three bins of 
environmental density and fit their relations, as before, with fixed slope 
(Table~\ref{fitPar}): high (Abell~496), moderate (Coma, Virgo, and Fornax 
Clusters, as well as the Shapley Supercluster), and low (groups).

Figure~\ref{DafeEnv} displays the residuals of the \afe-\sgm\ relation, 
$\delta$\afe, as a function of \sgm, for all low-mass galaxies with reliable 
environment information.  The red, green, and blue triangles represent the 
weighted-mean\footnote[1]{Because some references did not publish errors on 
\sgm, the mean \sgm\ is not weighted.} \afe\ and \sgm\ of the subsamples 
residing in environments of high, moderate, and low density, respectively; 
their corresponding zero points and intrinsic scatter are given by the dashed 
lines and colored bands. 

The majority of the low-mass ETGs in our study are from moderate-density 
environments.  Similar to the situation for the entire sample of low-mass 
objects, the moderate-density subsample has a zero point {\it and}\ intrinsic 
scatter consistent with those of the massive end.  This implies that the 
large intrinsic scatter of the low-mass sample comes mostly from objects in 
extreme environments. Relative to the high-mass end, low-mass objects in 
high-density environments have a higher zero point 
and larger scatter over 2\sgm\ significance; 
those in low-density environments 
are even more notable in terms of their lower zero point (5\sgm\ significance) 
and increased scatter (1\sgm\ significance). \\

\subsection{An Empirical Check} 
\label{R_index}

\begin{figure}
\epsscale{1.29}
\plotone{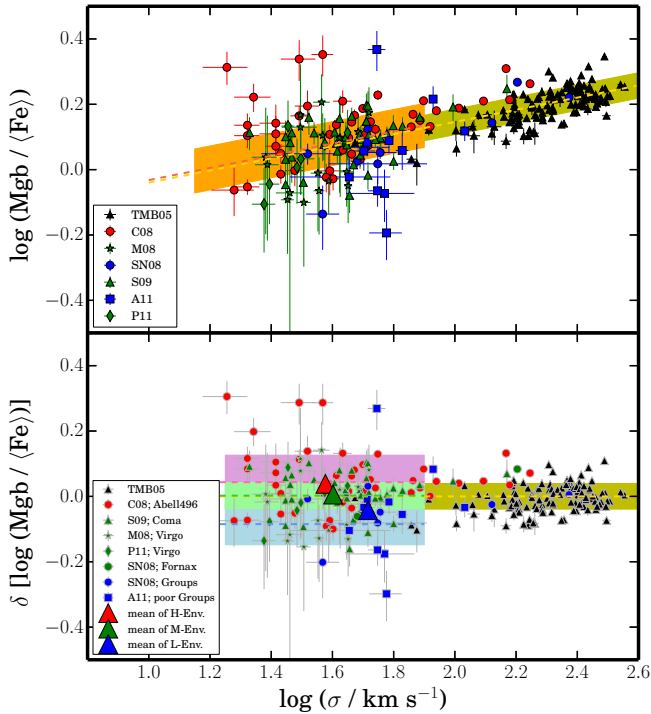}
\caption{ 
The upper and bottom panels are parallel plots to Figures~\ref{afeSgm} and \ref{DafeEnv}, respectively, with \afe\ replaced by \lgMF, illustrating that the relations are model-independent.  Symbols, lines and colors have the same meaning as in Figures~\ref{afeSgm} and \ref{DafeEnv}. \\
 \label{dIenv} }
\end{figure}
\begin{table}
\begin{center}
\caption{ The \lgMF-\sgm\ relation for different subsamples  
 \label{fitPar_I} }
\begin{tabular}{lccc}
\tableline\tableline
Group & $a$ & $b$ & $\epsilon_0$ \\
\tableline\tableline
High-\sgm\ & $0.186\pm0.016$ & $-0.226\pm0.037$ & $0.039\pm0.002$ \\
\tableline\tableline
Low-\sgm \\
\tableline\tableline
Total & $0.186$\footnotemark[1] & $-0.218\pm0.006$ & $0.067\pm0.008$ \\
H-Env & $0.186$\footnotemark[1] & $-0.182\pm0.014$ & $0.083\pm0.014$ \\
M-Env & $0.186$\footnotemark[1] & $-0.224\pm0.007$ & $0.040\pm0.007$ \\
L-Env & $0.186$\footnotemark[1] & $-0.311\pm0.025$ & $0.063\pm0.023$ \\
\tableline\tableline
\end{tabular}
\tablecomments{ \lgMF\ = $a \log \sigma  + b$. 
$\epsilon_0$ is the intrinsic scatter. 
H/M/L-Env represent the groups of low-mass 
ETGs with different levels of environmental density. \\
\footnotetext[1]{No error measurement, because the slope is fixed.} 
}
\end{center}
\end{table}

Because \afe\ is a model-dependent parameter, its values derived from 
different stellar population models may have systematic offsets. To mitigate 
these effects, we only use measurements derived from Lick indices, and 
we homogenize the derived \afe\ values to a common system (see \S 2).  
Nevertheless, it would be desirable to verify whether our results are 
robust with respect to model uncertainties.  
Independent of models, Mgb$/\langle{\rm Fe}\rangle$ is the most direct indicator of $\alpha$/Fe among the Lick indices.  Although this index ratio is also dependent on age and metallicity, it is most sensitive to $\alpha$/Fe.  
Seven of our literature sources published measurements of both Mgb and $\langle{\rm Fe}\rangle$, covering the full range of environmental density probed in this study.  
Figure~\ref{dIenv} presents the equivalent of the \afe-\sgm\ relation (Figures~\ref{afeSgm} and \ref{DafeEnv}), wherein we replace \afe\ with \lgMF.  
Table~\ref{fitPar_I} lists the fitting parameters of the \lgMF-\sgm\ relation for the different subsamples. 

All the main conclusions from \S 3.1 and \S 3.2 hold.  As before, massive 
ETGs follow a very tight relation between \lgMF\ and \sgm\ with relatively 
small intrinsic scatter, and, within the uncertainties, this relation 
can be extrapolated into the low-mass regime, but with larger intrinsic 
scatter.  When low-mass ETGs are grouped by their environments, those that 
reside in moderate-density clusters have a relation that is indistinguishable 
from that of massive ETGs in terms of zero point and intrinsic scatter, while 
those in the highest and lowest density environments have larger intrinsic 
scatter (3\sgm\ and 1\sgm\ significance) with higher and lower zero points 
(3\sgm\ significance), respectively.  This analysis, making use of a 
separate, model-independent estimate of \afe, confirm that the tight 
\afe-\sgm\ relation for massive ETGs can be extrapolated toward low masses, but 
that its intrinsic scatter and zero point differ in extreme environments. \\

\section{Discussion} 
\label{discuss} 

\subsection{The Origin of \afe-\sgm\ Relation} 
\label{D_aMR}

The mean \afe\ of galaxies arises from a complex interplay of many factors, 
including \tSF, IMF, stellar yields, SN~Ia explosion rate and delay time 
distribution, and gas inflow and outflow. 
The departure of the standard \afe-\sgm\ relation toward low masses, at least 
for ETGs residing in high- and low-density environments, implies that some of 
the physical drivers underlying this empirical scaling relation evidently 
change under these circumstances.  While it is difficult to assess the relative 
importance of the various potential factors, here we highlight the possible 
role of \tSF\ and SFH.

Two often-discussed candidate quenching mechanisms for massive ETGs are AGN 
feedback (e.g., Croton et al. 2006) and halo quenching (e.g., Dekel \& Birnboim 2006).  These quenching mechanisms appear to be mass-dependent, and more massive ETGs have shorter \tSF.
However, low-mass 
ETGs are quenched by different mechanisms. Some of them were satellite galaxies 
and were quenched by environmental processes, such as ram pressure stripping 
and strangulation, when they fell into the dark matter halos of nearby host 
galaxies (Gunn \& Gott 1972; Peng et al. 2015). Their \tSF\ might be more driven by environment than mass.
Moreover, galaxies with shallower potential wells are expected to experience 
even stronger environmental effects, which may result in shorter \tSF\ and 
higher \afe.  The effect of environment should be especially acute given that 
most low-mass ETGs sample reside in galaxy clusters.  Under 
these circumstances, we naively expect low-mass ETGs to exhibit a \afe-\sgm\ 
correlation with negative slope, opposite to that actually observed.  A 
positive \afe-\sgm\ correlation also contradicts the simplest expectations 
from internal quenching mechanisms such as radiative and energetic feedback 
from SNe and high-mass stars, which should operate more effectively 
and stop star formation earlier (thus boosting \afe) in lower mass galaxies.

The \afe-\sgm\ relation may also be a consequence of SFH.  From cosmological 
semi-analytical simulations, De Lucia et al. (2006) showed that more 
massive ETGs had more peaked SFR distributions, which reached their peak 
at higher redshifts. However, they only studied massive galaxies with 
stellar masses larger than $4\times10^9\,M_\odot$. Moreover, other studies
have not been able to reproduce the positive correlation between \afe\ and 
\sgm\ for massive ETGs without including special recipes (e.g., Arrigoni et 
al. 2010; Yates et al. 2013; Segers et al. 2016).  Segers et al. (2016) 
extended their simulations to galaxies of lower masses and reproduced a 
positive \afe-\sgm\ correlation for stellar masses between $10^8\,M_\odot$ 
and $10^{10}\,M_\odot$. It is unclear, however, whether these simulations 
apply to the situation at hand.  Segers et al. only considered central 
galaxies that are still forming stars at $z \approx 0$. \\

\subsection{The Role of Environment} 
\label{D_env} 


We find that low-mass ETGs from the densest environments are, on average, 
offset toward larger \afe.  A plausible implication of this finding is that 
high environmental density induces short \tSF. As discussed in 
Section~\ref{D_aMR}, low-mass galaxies in high-density regions might 
experience environmental quenching earlier, which would systematically elevate 
their \afe.  At the same time, environmental density also influences early 
star formation processes in low-mass galaxies. Santos et al. (2015) found that 
star-forming galaxies in the central regions of high-redshift clusters have 
higher SFRs.  This suggests that present-day low-mass ETGs residing in denser 
environments may have experienced more intense star formation at early times, 
resulting in their higher observed \afe. 
By contrast, the average \afe\ of low-mass ETGs is lower in galaxy groups, 
indicating more extended SFHs of low-mass galaxies in low-density environments. 
This is consistent with the findings of Geha et al. (2012). All the isolated 
low-mass galaxies without H$\alpha$ emission in their sample showed evidence
of recent starbursts, implying bursty SFHs over an extended period.

In both of the extreme (high- and low-density) environments highlighted in 
this work, the large scatter of \afe\ at low-mass can be explained by 
stochastic chemical evolution processes in non-equilibrium systems. According 
to the main sequence of star-forming galaxies (e.g., Wuyts et al. 2011), 
low-mass galaxies have low SFRs. Under such conditions, massive stars and 
SNe~II appear stochastically because the number of forming stars is not large 
enough to sample the IMF completely (e.g., Lee et al. 2009; Fumagalli et al. 
2011).  Furthermore, some SNe in the early universe may produce special 
abundance patterns, such as both low \afe\ and low [Fe/H] (Kobayashi et al. 
2014; Simon et al. 2015).   The dispersion in \afe\ also reflects the details 
of the gas evolution after the appearance of the first generation of stars.  
For example, star formation may cease in hot bubbles after the surrounding 
neutral gas is ionized, but it may restart in colder regions afterwards (e.g., 
Sobral et al. 2015).  Outflows with high \afe\ ejected by these stars, if 
still bound to the halo, might fall back and contribute to subsequent star 
formation.  Both scenarios would temporally increase the mean galactic \afe. 

Yet, it is curious that the large scatter in \afe\ is not seen in all low-mass 
ETGs, but rather only in the subset residing in extremely high- or low-density 
environments.  After all, the intrinsic scatter of the \afe-\sgm\ relation 
over the same low-mass range in normal galaxy clusters is as tight as in 
high-mass ETGs.  We have no explanation for this.

\section{Summary}
\label{summary}

We assemble 708 nearby ETGs from 15 literature data sets to investigate the 
\afe-\sgm\ relation across a wide range of galactic mass and environment. Our 
sample spans $\sigma \approx 18 - 360$ \kms, with environments covering both 
galaxy clusters and groups,
among them 192 low-mass ETGs with $\sigma < 80$~\kms. 

We recover the well-known tight \afe-\sgm\ relation among massive 
($\sigma\gtrsim80$~\kms) ETGs.  The relation extends into the low-mass 
regime, especially for low-mass ETGs residing in moderate-density galaxy 
clusters and superclusters, which maintain the same zero point and intrinsic 
scatter defined by high-mass systems.  By contrast, the zero point of the 
\afe-\sgm\ relation for low-mass ETGs is higher in the densest cluster and 
suppressed in the lower density groups; the intrinsic scatter of the relation 
is strikingly larger in both of these extreme environments.

We suggest that both mass and environment regulate \afe\ in low-mass ETGs. 
While \afe\ is governed by total galactic mass in normal galaxy clusters, it 
is elevated by the earlier quenching in the densest environments and suppressed
in due to the more extended SFHs in galaxy groups.  The overall low SFR of 
low-mass systems induces stochasticity in their chemical enrichment history, 
which can plausibly account for the observed increased scatter in \afe. \\

\acknowledgments
We thank an anonymous referee for comments that helped improve our paper.  
YQL is grateful for the great academic environment at KIAA; she also appreciates discussions with Yingjie Peng and Andrew McWilliam, as well as the help from Minjin Kim with the fitting algorithms.  This work was supported by grant 2016YFA0400702 from the Ministry of Science and Technology of China.  We also acknowledge support from the National Natural Science Foundation of China under grant No.\ 11573002 and 11473002, and from the Strategic Priority Research Program, ``The Emergence of Cosmological Structures'', of the Chinese Academy of Sciences, under grant No. XDB09030102 and XDB09000105. \\


\end{document}